\documentclass{PoS}

\title{Precise Prediction of the Dark Matter 
Relic Density within the MSSM}

\ShortTitle{Precise Prediction of the Dark Matter 
Relic Density within the MSSM}

\author{\speaker{Julia Harz}\\
        Sorbonne Universit\'es, Institut Lagrange de Paris (ILP), 98 bis Boulevard Arago, 75014 Paris, France\\
Sorbonne Universit\'es, UPMC Univ Paris 06, UMR 7589, LPTHE, F-75005, Paris, France\\
CNRS, UMR 7589, LPTHE, F-75005, Paris, France \\
Department of Physics and Astronomy, University College London, London WC1E 6BT, United Kingdom\\
        E-mail: \email{jharz@lpthe.jussieu.fr}}

\author{Bj\"orn Herrmann\\
LAPTh, Universit\'e Savoie Mont Blanc, CNRS, 9 Chemin de Bellevue, F-74941 Annecy-le-Vieux,
France}

\author{Michael Klasen, Karol Kova\v{r}\'ik, Patrick Steppeler\\
Institut f\"ur Theoretische Physik, Westf\"alische Wilhelms-Universit\"at M\"unster, Wilhelm-Klemm-Stra{\ss}e 9, D-48149 M\"unster, Germany}

\abstract{With the latest Planck results the dark matter relic density is determined to an unprecedented precision. In order to reduce current theoretical uncertainties in the dark matter relic density prediction, we have calculated next-to-leading order SUSY-QCD corrections to neutralino (co)annihilation processes including Coulomb enhancement effects. We demonstrate that these corrections can have significant impact on the cosmologically favoured MSSM parameter space and are thus of general interest for parameter studies and global fits.}

\FullConference{The European Physical Society Conference on High Energy Physics\\
		22--29 July 2015\\
		Vienna, Austria}

\begin{document}

\section{Introduction}
Different observations point us to the existence of physics beyond the Standard Model (SM). For example, there are ample hints for the existence of dark matter, quantified by the measurement of the dark matter relic density to an unprecedented precision by the Planck collaboration \cite{Planck}
\begin{equation}
	\Omega_{\mathrm{CDM}}h^2 = 0.1198 \pm 0.0015.
	\label{Planck}
\end{equation}
This quantity, $\Omega_{\chi} h^2$, can be theoretically determined as $\Omega_{\chi} h^2 ~=~ m_{\chi} n_{\chi} / \rho_{\rm crit}$, with $\rho_{\rm crit}$ being the critical density of the Universe and $\chi$ denoting the dark matter candidate with $m_{\chi}$ being the corresponding mass. Its today's number density $n_{\chi}$ can be calculated via the Boltzmann equation
\begin{equation}
	\frac{\mathrm{d}n_\chi}{\mathrm{d}t} = -3 H n_\chi 
		- \left\langle\sigma_{\mathrm{ann}}v\right\rangle \Big[ n_\chi^2 
		- \left( n_\chi^{\mathrm{eq}} \right)^2 \Big],
	\label{Eq:Boltzmann}
\end{equation}
where $H$ is the Hubble parameter and $\sigma_{\mathrm{ann}}$ the effective (co)annihilation cross section. This effective cross section takes into account all possible processes of (co)annihilating $Z_2$-odd particles $i,j$ into SM particles given as
\begin{equation}
	\left\langle \sigma_{\mathrm{ann}}v\right \rangle = 
		\sum_{i,j} \langle \sigma_{ij}v_{ij} \rangle \frac{n_i^{\mathrm{eq}}}{n_\chi^{\mathrm{eq}}}
			\frac{n_j^{\mathrm{eq}}}{n_\chi^{\mathrm{eq}}},
	\label{Eq:Sigma}
\end{equation}
where $\sigma_{ij}$ is the cross section of two (co)annihilating $Z_2$-odd particles and $v_{ij}$ the corresponding relative velocity. The number density ratios can be written as $n_{i,j}^{\mathrm{eq}} / n_\chi^{\mathrm{eq}} ~\sim~ \exp\left\{ - (m_{i,j}-m_\chi) / T \right\}$.
This indicates that a process will be enhanced when the particle $i$ (or $j$) is almost degenerate in mass with the dark matter particle $\chi$. Thus, not only the standard annihilation of two dark matter particles is of interest but also processes involving particles close in mass to the dark matter particle.
Based on this calculation, it is possible to calculate the dark matter relic density for different parameter sets and to use this quantity to set constraints on the MSSM parameter space. This is generally done in global fits and parameter studies (e.g. \cite{deVries:2015hva, Fowlie:2011mb}). However, these calculations are usually based on programs like {\tt micrOMEGAs} \cite{micrOMEGAs} or {\tt DarkSUSY} \cite{DarkSUSY}, which calculate the relic density based on an (effective) tree-level calculation of $\sigma_{ij}$, without including higher-order corrections or Coulomb enhancement effects. To study the impact of those higher order corrections and to achieve a more precise theoretical prediction of the dark matter relic density, we calculate SUSY-QCD next-to-leading order (NLO) corrections to the processes involved.
\begin{figure}[t]
\centering
\includegraphics[clip,width=0.75\linewidth]{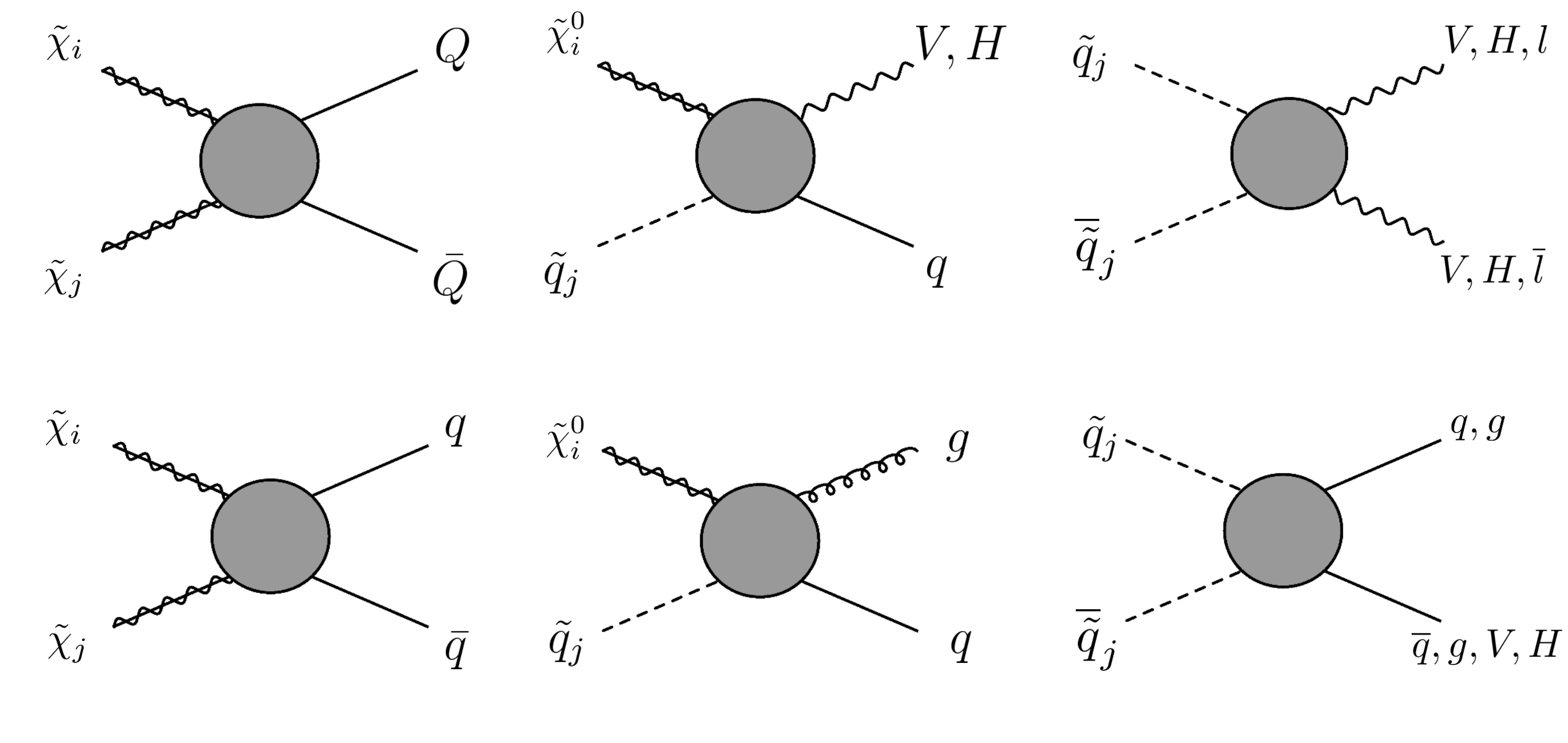}
\caption{Overview of the main classes of diagrams relevant for SUSY-QCD corrections to (co)annihilation processes: gaugino (co)annihilation (left column), neutralino-stop coannihilation (middle column), stop-antistop annihilation (right column). The blob indicates all possible tree- and one-loop level diagrams. }
\label{Fig:DMNLOoverview}  
\end{figure}
Fig.~\ref{Fig:DMNLOoverview} shows an overview of the main relevant classes of processes: (co)annihilation of gauginos (left column), coannihilation of neutralino-stop processes (middle column) and stop-antistop annihilation (right column). Except of stop-antistop annihilation into coloured final states, all processes have been calculated \cite{DMNLO_AFunnel,DMNLO_mSUGRA, DMNLO_NUHM, DMNLO_ChiChi, DMNLO_Stop1, DMNLO_Stop1b, DMNLO_Stop1c, DMNLO_Stop2, DMNLO_Stopstop} and their impact on the relic density discussed. In the following, we would like to highlight some of the crucial findings for each of those processes.

\section{Technical Details}
As a full discussion of technical details would go beyond the scope of this proceeding, we refer the interested reader to our previous publications \cite{DMNLO_AFunnel,DMNLO_mSUGRA, DMNLO_NUHM, DMNLO_ChiChi, DMNLO_Stop1, DMNLO_Stop1b, DMNLO_Stop1c, DMNLO_Stop2, DMNLO_Stopstop} for a detailed discussion and mention mainly the key features of our calculation in a short manner in the following. 

For renormalising the appearing ultraviolet (UV) divergences we use a hybrid on-shell /  $\overline{\mathrm{DR}}$ scheme. We choose the parameters  $A_t, A_b, m_{\tilde{t}_1}, m_{\tilde{b}_1}, m_{\tilde{b}_2}$ along with the heavy quark masses $m_t, m_b$ as input parameters and define $m_b, A_t, A_b$ in the $\overline{\mathrm{DR}}$ scheme, while treating the remaining ones ($m_t, m_{\tilde{t}_1}, m_{\tilde{b}_1}, m_{\tilde{b}_2}$) in the on-shell scheme. This allows for a stable calculation of all annihilation and coannihilation processes over a wide parameter range. The renormalisation and factorisation scale are set to 1~TeV that corresponds to the scale at which the $\overline{\mathrm{DR}}$-input values are given. 
Further, we use effective Yukawa couplings for the bottom quark including QCD corrections up to $\mathcal{O}(\alpha_s^4 )$ as well as top-quark induced corrections. As in the MSSM, sizeable corrections can appear for large $\tan \beta$ or large $A_b$, these effects have been resummed to all orders and been taken into account as well. For further details we refer to \cite{DMNLO_ChiChi,DMNLO_Stop1} and references therein. For the evolution of the bottom quark mass from $m_b^{\mathrm{SM}, \overline{\mathrm{MS}}}(m_b)$ to $m_b^{\mathrm{MSSM}, \overline{\mathrm{DR}}}(Q)$, we refer to \cite{DMNLO_ChiChi} and corresponding references. Regarding the renormalisation group evolution for $\alpha_s$ from $\alpha_s^{\overline{\mathrm{MS}},\mathrm{SM},n_q=5}(m_Z^2)$ to $\alpha_s^{\overline{\mathrm{DR}},\mathrm{MSSM},n_q=6}(Q^2)$ we refer again to \cite{DMNLO_Stop1} and references therein.
For the treatment of the infrared (IR) divergences, again regularised dimensionally, two main ansaetze exist: Dipole subtraction and phase space slicing. For stop-antistop annihilation and neutralino-stop coannihilation, we use a one (two) cutoff slicing method \cite{HarrisOwens}, for the gaugino (co)annihilation we use the Catani-Seymour subtraction \cite{Catani1}. For further details on the specific treatment and other subtleties like the treatment of on-shell states, we refer again to our corresponding papers.

\section{Gaugino (Co)annihilation}
\begin{figure*}
\centering
	\includegraphics[keepaspectratio=true,trim = 5mm 20mm 5mm 18mm, clip=true, width=0.85\textwidth]{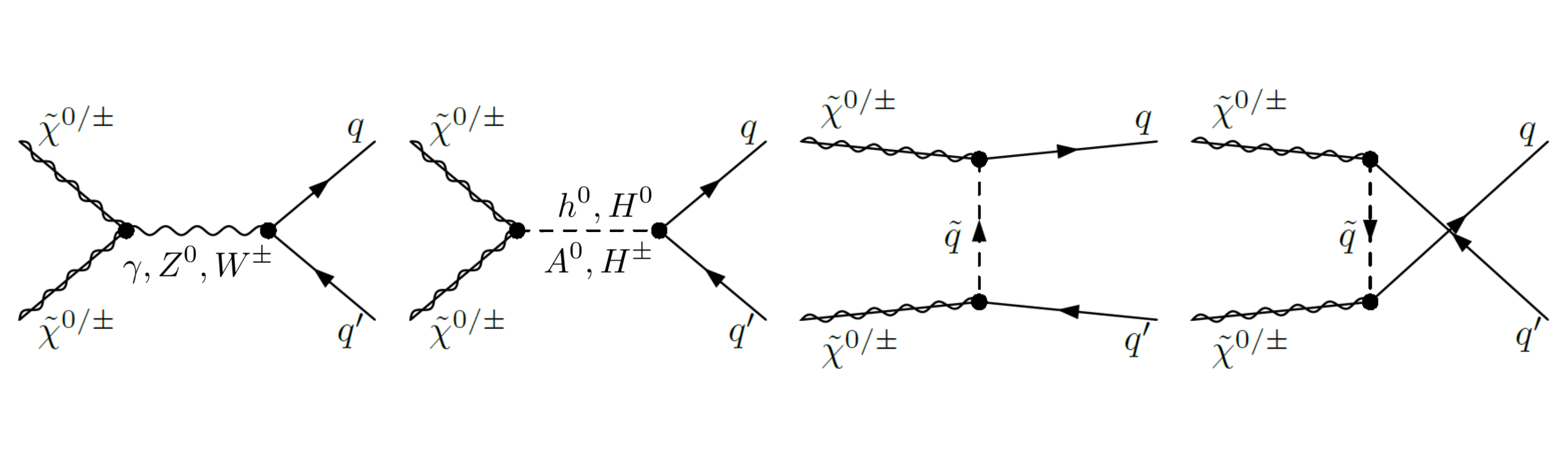}
	\caption{Tree-level diagrams contributing to gaugino (co)annihilation.}
	\label{Fig:Gaugino_treelevel}
\end{figure*}

First, we discuss a scenario with dominant gaugino (co)annihilation in a pMSSM-11 Scenario (A) with parameters as listed in Tab.~\ref{Tab:Scenarios}. It fulfils the current experimental bounds on the lightest Higgs boson mass, the constraints from the branching ratio of $b\rightarrow s\gamma$, and the relic density within theoretical uncertainties. The generic diagrams contributing at tree level are depicted in Fig.~\ref{Fig:Gaugino_treelevel}.%
\begin{table*}
		\begin{center}
		\footnotesize
		\begin{tabular}{|c|ccccccccccc|}
			\hline 
			  &  ~$\tan\beta$~ & ~~$\mu$~~ & ~~$m_A$~~ & ~~$M_1$~~ & ~~$M_2$~~ & ~~$M_3$~~ & ~~$M_{\tilde{q}_{1,2}}$~~ & ~~$M_{\tilde{q}_3}$~~ & ~~$M_{\tilde{u}_3}$~~ & ~~$M_{\tilde{\ell}}$~~ & ~~$T_t$~~  \\
			\hline
			A & 13.4 & 1286.3 & 1592.9 & 731.0 & 766.0 & 1906.3 & 3252.6 & 1634.3 & 1054.4 & 3589.6 & -2792.3\\
			B  & 5.8 & 2925.8 & 948.8 & 335.0 & 1954.1 & 1945.6 & 3215.1 & 1578.0 & 609.2 & 3263.9 & 3033.7 \\  
			C  & 16.3 & 2653.1 & 1917.9 & 750.0 & 1944.1 & 5832.4 & 3054.3 & 2143.7 & 1979.0 & 2248.3 & -3684.1 \\
			\hline
		\end{tabular}
		\end{center}
		\caption{Parameters defining the example scenario in the pMSSM-11. All quantities except $\tan\beta$ are given in GeV.}	
		\label{Tab:Scenarios}
\end{table*}
In this scenario, $\tilde{\chi}^0_1 \tilde{\chi}^{\pm}_1 \to t\bar{b}$ is the dominant contribution with $43~\%$. With $m_{\tilde{\chi}^0_1}$ and $m_{\tilde{\chi}^\pm_1}$ being close in mass (c.f. Tab.~\ref{Tab:ScenarioMasses}), the Boltzmann suppression is less stringent. Further, the process is with respect to the usual neutralino pair annihilation kinematically favoured due to a charged Higgs resonance ($m_{\tilde{\chi}^0_1} + m_{\tilde{\chi}^\pm_1} \approx m_{H^\pm}$).
\begin{table*}
\footnotesize
	\begin{tabular}{|c|cccc|cc|c|cc|}
		\hline
			 & ~~$m_{\tilde{\chi}^0_1}$~~ & ~~$m_{\tilde{\chi}^0_2}$~~ & ~~$m_{\tilde{\chi}^0_3}$~~ & ~~$m_{\tilde{\chi}^0_4}$~~ & ~~$m_{\tilde{\chi}^{\pm}_1}$~~ & ~~$m_{\tilde{\chi}^{\pm}_2}$~~ & ~~$m_{h^0}$~~ & ~~$\mathrm{BR}(b\rightarrow s\gamma)$~~ &~~$\Omega_{\tilde{\chi}^0_1} h^2$~~  \\
			\hline 
			A & 738.2 & 802.4 & 1288.4 & 1294.5 & 802.3 & 1295.1  & 126.3 & $3.0\cdot 10^{-4}$ & 0.1243 \\			
			\hline		
	\end{tabular}
	\vspace{0.3cm}
	
		\begin{tabular}{|c|cc|c|cc|}
			\hline 
			& ~~$m_{\tilde{\chi}^0_1}$~~ & ~~$m_{\tilde{t}_1}$~~&~~$m_{h^0}$~~&~~${\rm BR}(b\to s\gamma)$ ~~&~~$\Omega_{\tilde{\chi}_1^0}h^2$\\
			\hline
			B & 338.3  & 375.6 & 122.0   & 0.1136 & $3.25 \cdot 10^{-4}$ \\
			\hline
		\end{tabular}
		
	\vspace{0.3cm}		
	\begin{tabular}{|c|cc|cccc|ccc|cc|}
		\hline
			 & $m_{\tilde{\chi}^0_1}$ & $m_{\tilde{t}_1}$ & $m_{\tilde{t}_2}$&$m_{\tilde{b}_1}$&$m_{\tilde{\chi}^0_2}$ & $m_{\tilde{\chi}^{\pm}_1}$ & $m_{h^0 }$ & $m_{H^0 }$& $m_{H^{\pm}}$ & $\mathrm{BR}(b\rightarrow s\gamma)$ & $\Omega_{\tilde{\chi}^0_1} h^2$\\
			\hline 
			C & 758.0 & 826.1 & 1435.1 & 1260.5 & 1986.7 & 1986.8 & 128.8 & 1917.4  & 1919.6 &  $3.1\cdot10^{-4}$ & 0.1146\\		
			\hline		
	\end{tabular}
		\caption[]{Sparticle masses, Higgs mass(es) and selected observables of the reference scenarios of Tab.\ \ref{Tab:Scenarios}. All masses are given in GeV.}

	\label{Tab:ScenarioMasses}
\end{table*}
With a large enough $\tan \beta = 13.4$, the bottom Yukawa coupling is enhanced, thus favouring the $t\overline{b}$ final state. A similar argument holds for the process $\tilde{\chi}^0_1 \tilde{\chi}^0_2 \to t\bar{b}$, which is less Boltzmann suppressed due to the $\tilde{\chi}_2^0$ being close in mass to the LSP, and favoured due to a resonant pseudo-scalar Higgs boson ($m_{\tilde{\chi}^0_1} + m_{\tilde{\chi}^0_2} \approx m_{A^0}$), It thus contributes with $23~\%$. The standard pair annihilation amounts to $9.1\%$, followed by pair annhilation of $\tilde{\chi}^0_2$ and $\tilde{\chi}^{\pm}_1$, which are highly suppressed by the large mass difference to the lightest supersymmetric particle (LSP).

\begin{figure*}
	\includegraphics[width=0.53\textwidth]{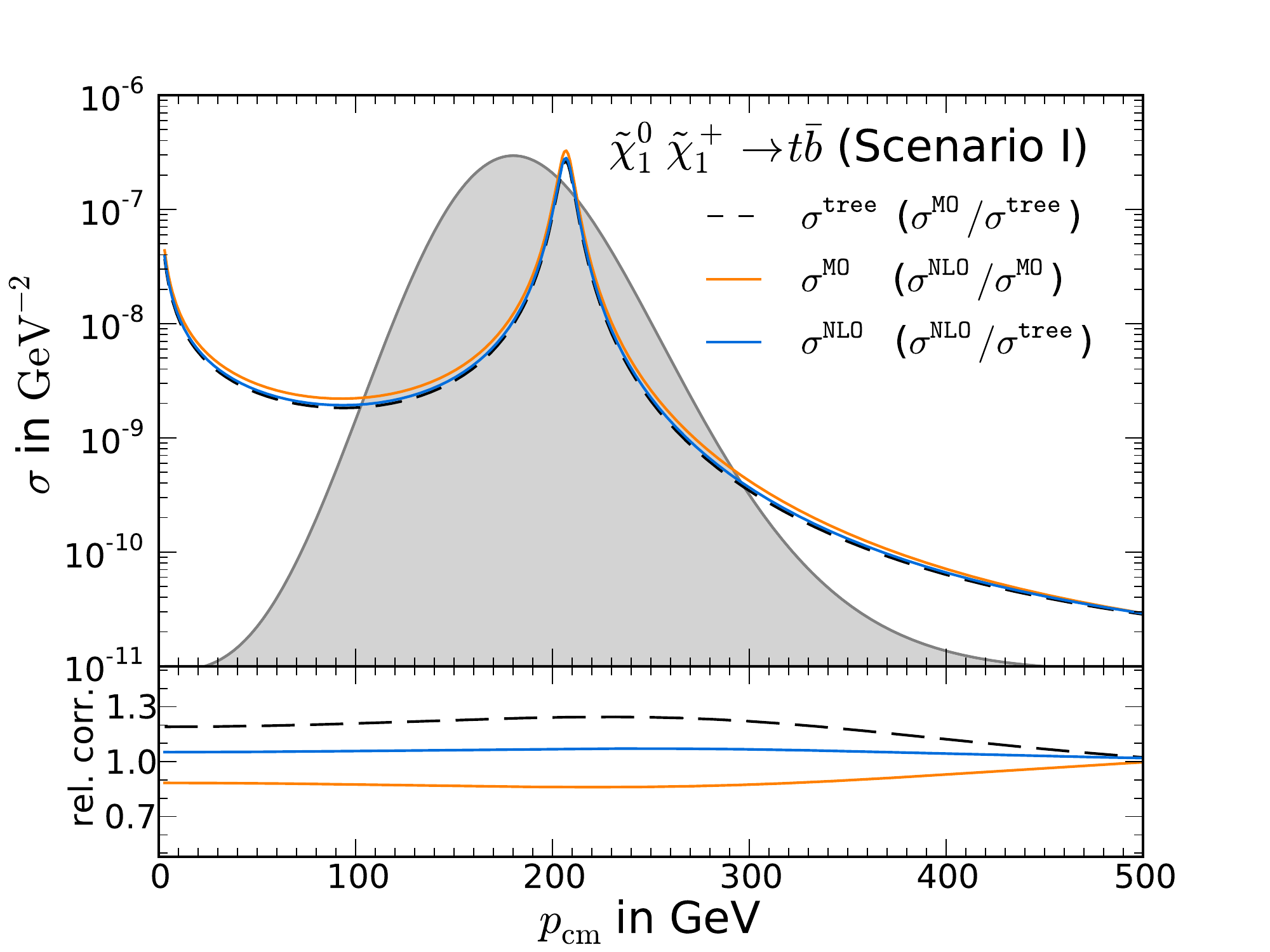}
	\includegraphics[width=0.53\textwidth]{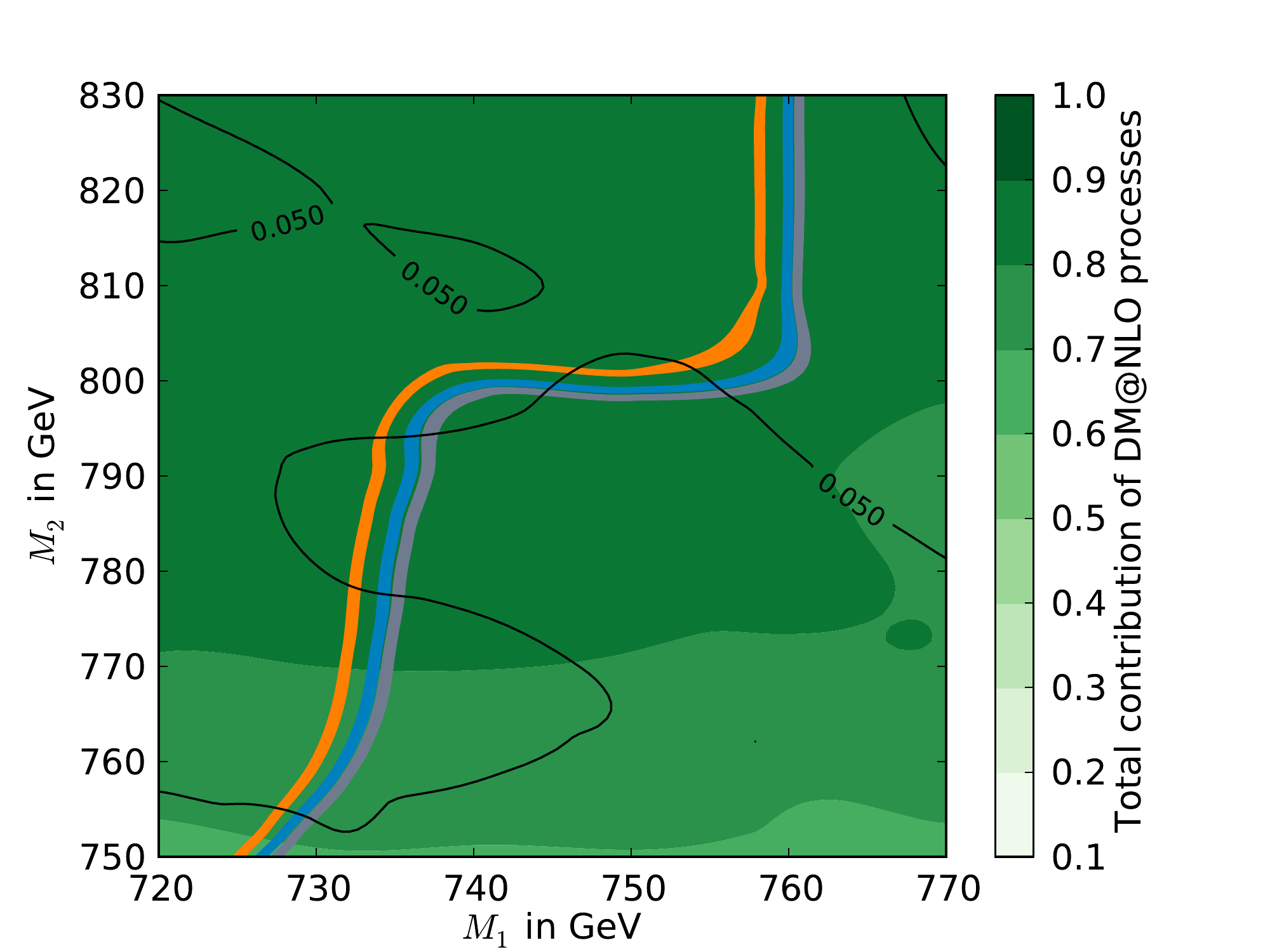}
	\caption{Left: Cross section calculated based on a tree level calculation (black dashed line), at NLO (blue solid line) and with {\tt MO} (orange solid line). The grey area indicates the thermal distribution in arbitrary units. The lower panel shows different ratios as indicated in the legend. Right: $M_1$-$M_2$-plane showing the region of parameter space compatible with the 1-$\sigma$ Planck limit based on our tree-level calculation (grey), NLO (blue) and {\tt MO} (orange). The black contours indicate the relative shift between the NLO and LO calculation.}
	\label{Fig:GauginoResults}
\end{figure*}

We have calculated the full SUSY-QCD NLO corrections to the diagrams shown in Fig.~\ref{Fig:Gaugino_treelevel} and refer to \cite{DMNLO_ChiChi} for an overview. The results are exemplarily shown in Fig.~\ref{Fig:GauginoResults}. On the left hand side, the cross section in dependence of the center-of-mass momentum $p_{\mathrm{cm}}$ for the process $\tilde{\chi}^0_1 \tilde{\chi}^{\pm}_1 \to t\bar{b}$ based on different calculations is shown. In orange, the default value of {\tt micrOMEGAs} ({\tt MO}) is depicted, in  black dashed our tree-level calculation and in blue our NLO calculation. The grey area indicates the velocity distribution in arbitrary units. The peak at around $p_{\mathrm{cm}} \approx 200~\mathrm{GeV}$ arises from the aforementioned $H^+$ resonance. Due to a different treatment of the bottom quark mass, our tree-level calculation and the {\tt micrOMEGAs} result differ around $20~\%$. Our loop calculation with respect to the result of {\tt micrOMEGAs}, however, deviate only about $10-15~\%$. This can be explained due to the fact that {\tt MO} already includes effective bottom Yukawa couplings.

On the right hand side of Fig.~\ref{Fig:GauginoResults}, the $M_1-M_2$-plane around the example scenario is shown. The coloured bands (orange {\tt MO}, grey LO, blue NLO) indicate the region of parameter space which is compatible with the 1-$\sigma$ limit by Planck. Including SUSY-QCD corrections at one loop, the relic density band is shifted by around 5\% with respect to the tree level calculation and around 10\% with respect to the {\tt MO} band. This shift is larger than the experimental uncertainty by the Planck measurement.

\section{Neutralino-Stop Coannihilation}
\begin{figure*}
	\includegraphics[width=0.49\textwidth]{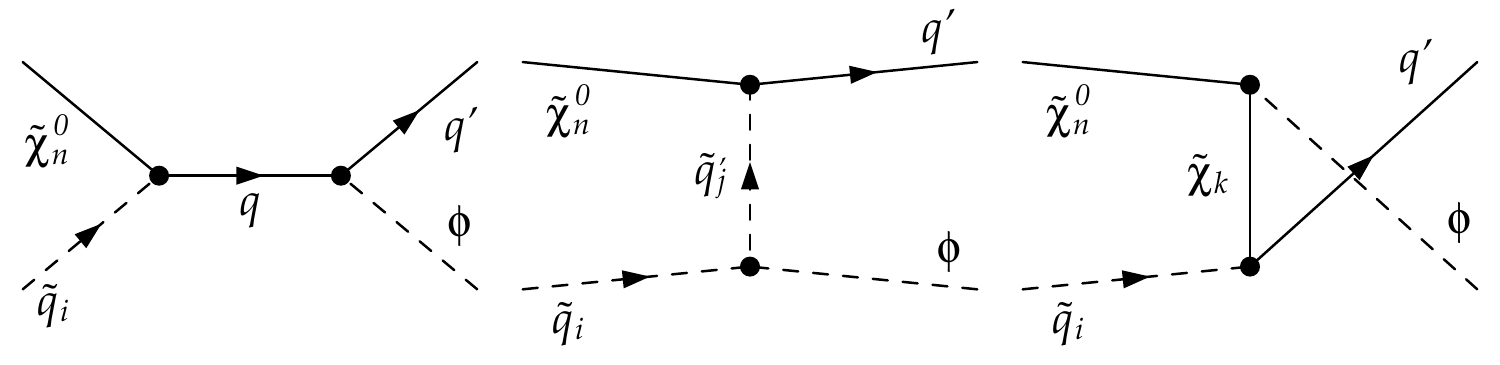}
	\includegraphics[width=0.49\textwidth]{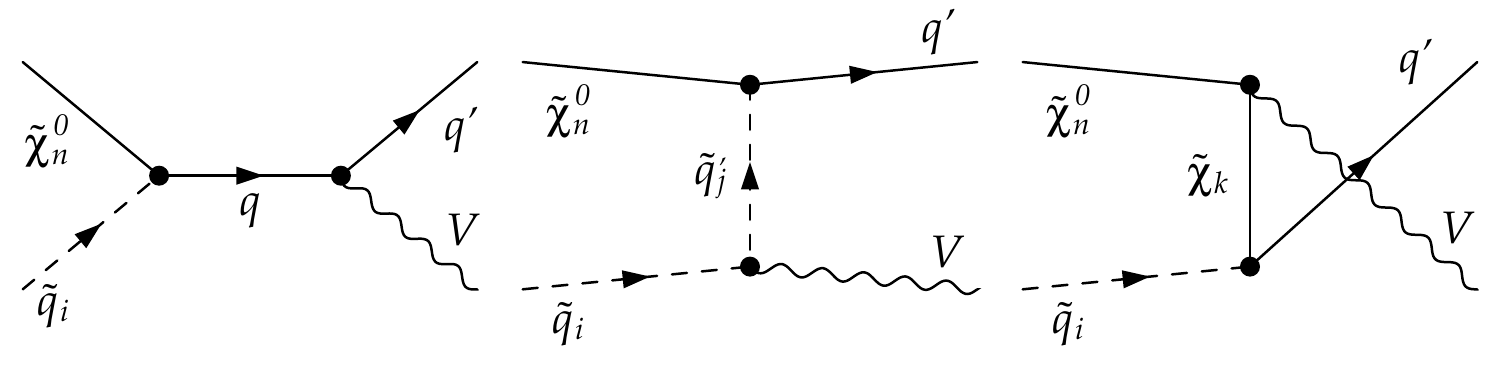}
	\caption{Tree level diagrams for neutralino-squark coannihilation into a quark and a Higgs boson ($\phi = h^0, H^0, A^0, H^\pm$) and a vector boson ($V =  \gamma, g, Z^0, W^\pm$). The $u$-channel diagram is absent for a photon or gluon in the final state.}
	\label{Fig:StopCoanni_treelevel}
\end{figure*}

As a second example, we focus on a scenario (B) with dominant neutralino-stop coannihilation. The parameters within the pMSSM-11 are given in Tab.~\ref{Tab:Scenarios}. The corresponding masses and observables are listed in Tab.~\ref{Tab:ScenarioMasses}, in Fig.~\ref{Fig:StopCoanni_treelevel} the leading-order diagrams are shown. This scenario is in particular interesting as it features a large trilinear coupling. This favours, on the one hand, a large stop mass splitting and makes, thus, the lightest stop accessible for coannihilation with the LSP. On the other hand, it helps for enhancing the Higgs mass to the observed value via the stop loop contribution to the Higgs mass. As such a scenario is difficult to access at the LHC, it is still a viable and interesting region of parameter space where the MSSM could hide. With the lightest stop and the lightest neutralino being close in mass, the Boltzmann suppression is less strong and this type of process can thus dominate the (co)annihilation cross section $\sigma_{\mathrm{ann}}$ with $61\%$ in Scenario B. It is mainly dominated by the $tg$ and $th$ final state with each contributing $23\%$, followed by $10\%$ of $b W^\pm$ and $5\%$ of $tZ^0$ final state. With additional $15\%$ contribution of $\tilde{\chi}_1^0 \tilde{\chi}_1^0 \to t \bar{t}$, we correct in total $76\%$ of all processes contributing to the relic abundance.

\begin{figure*}
	\begin{center}
		\includegraphics[scale=0.36]{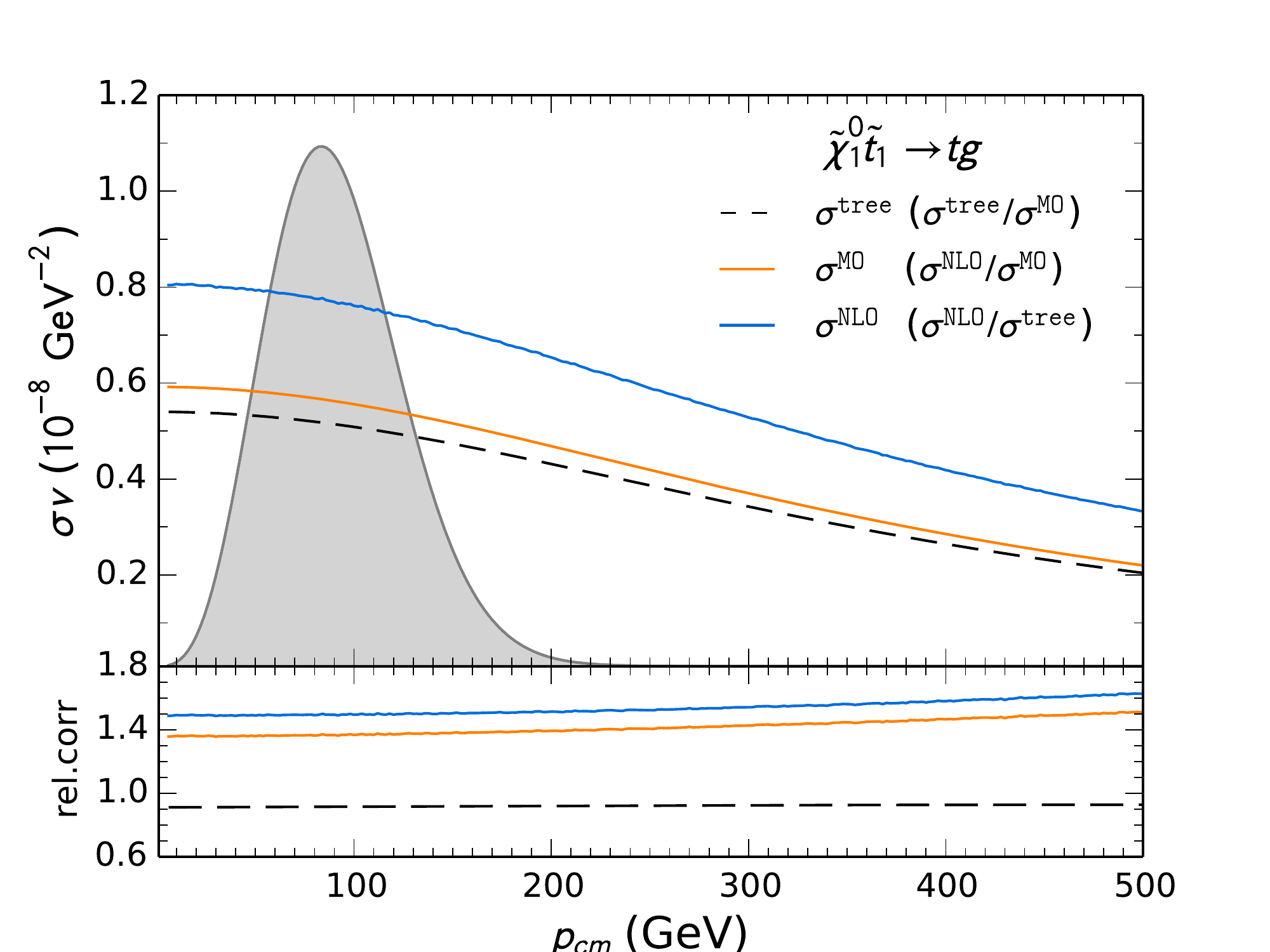}
		\includegraphics[scale=0.36]{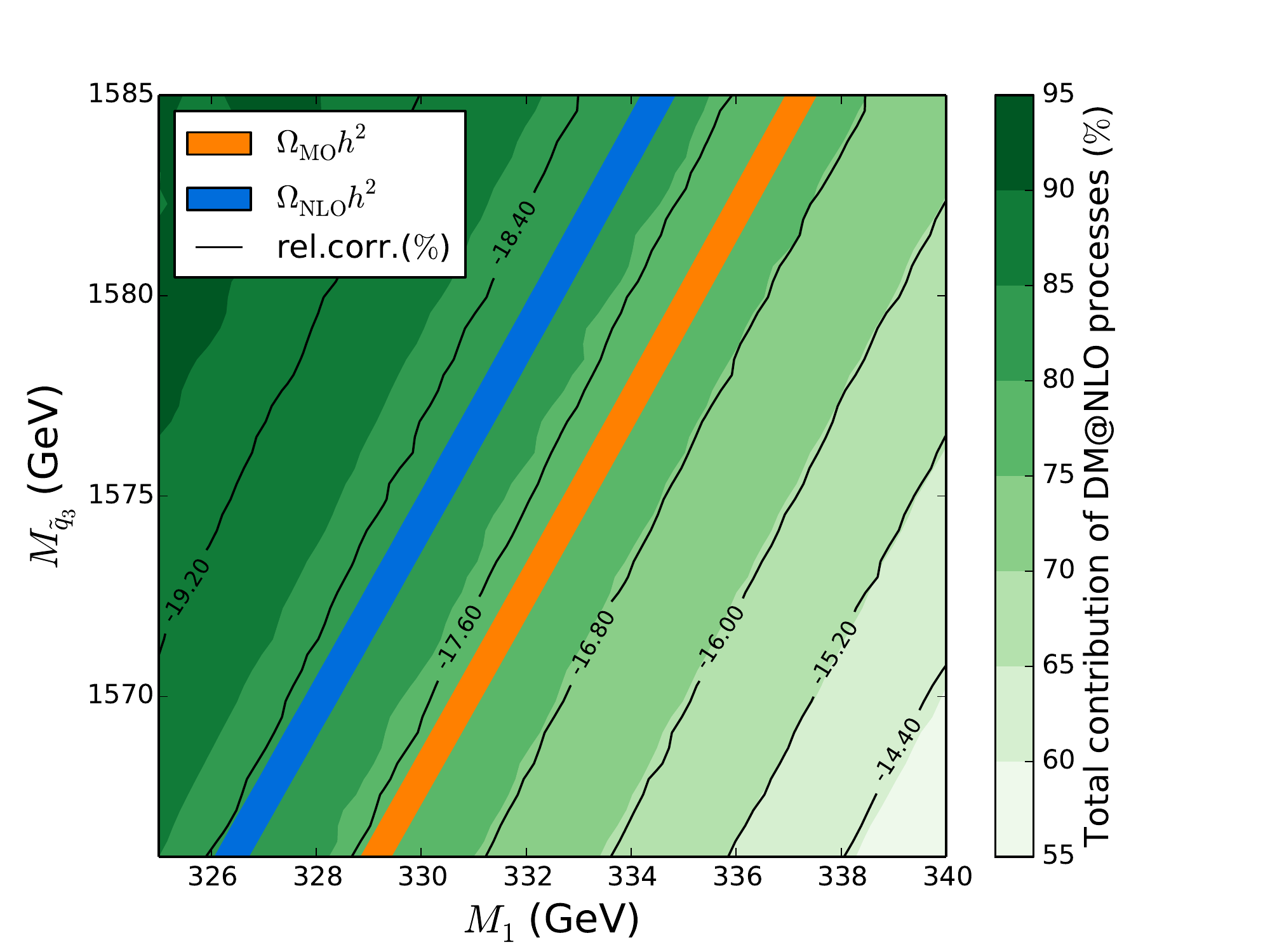}
	\end{center}
	\caption{Left: Tree-level (black dashed line), full one-loop (blue solid line) and {\tt micrOMEGAs} (orange solid line) cross sections. The absolute value of $\sigma v$ together with the thermal velocity distribution (in arbitrary units evaluated at the freeze-out temperature) is depicted in the upper part, whereas in the lower part the corresponding relative corrections (second item in the legend) are shown. Right: $M_1$--$M_{\tilde{q}_3}$ plane in the vicinity of scenario B. The region of parameter space compatible with the Planck results is shown as coloured bands. Blue indicates the relic abundance based on our full NLO calculation, orange based on {\tt MO}.}
	\label{Fig:CoanniResults}
\end{figure*}

We have calculated the full SUSY-QCD NLO corrections to all processes depicted in Fig.~\ref{Fig:StopCoanni_treelevel} (for details we refer to \cite{DMNLO_Stop1, DMNLO_Stop2}). The impact of the corrections are exemplarily shown in Fig.~\ref{Fig:CoanniResults}.
The plot on the left hand side shows the effect of the NLO corrections for the process $\tilde{\chi}_1^0 \to tg$. Our performed tree-level calculation deviates less than $10\%$ from the {\tt MO} result. This existing deviation is well understood and can be explained by the chosen renormalisation scheme. However, a significant shift of around $40\%$ of our NLO calculation with respect to {\tt MO} is observed. This is due to NLO corrections which involve $\mathcal{O}(\alpha_s^2)$ contributions in case of the $tg$ final state, naturally leading to large NLO corrections. The combined impact of all corrected channels is demonstrated in the right plot of Fig.~\ref{Fig:CoanniResults}. It shows the 1-$\sigma$ Planck band for the default {\tt MO} calculation (orange) and for the NLO calculation (blue). A clear shift of the NLO band with respect to {\tt MO} is visible, leading to a relative correction of up to $18~\%$. Indicated by green colour, around $80~\%$ of all contributing (co)annihilating channels are corrected. The third class of processes in this context is stop-antistop annihilation, which will be discussed in the subsequent section.

\section{Stop-antistop Annihilation}
\begin{figure*}
\begin{minipage}{0.6\textwidth}
	\includegraphics[keepaspectratio=true,trim = 35mm 115mm -15mm 100mm, clip=true, width=1.0\textwidth]{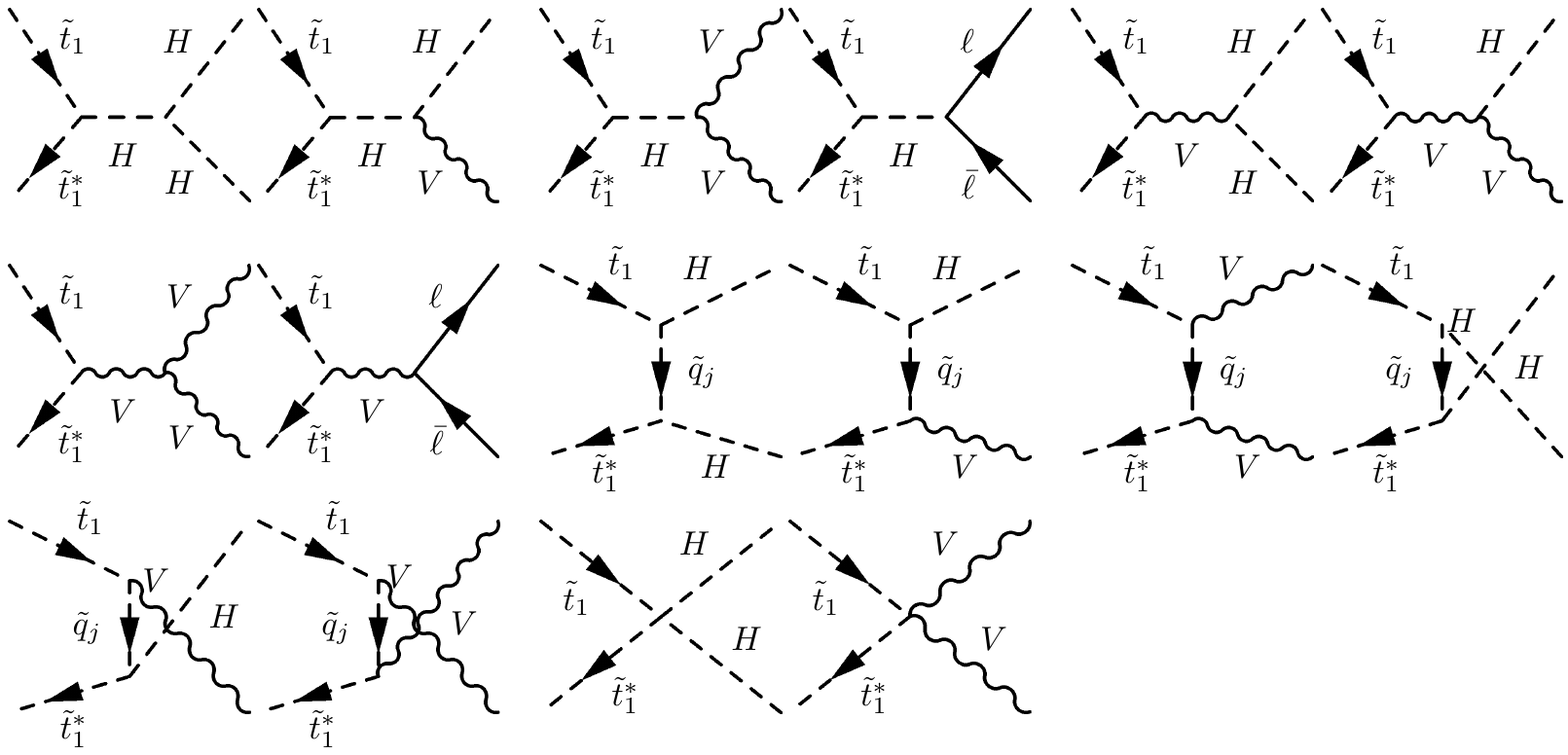}
\end{minipage}
\hfill
\begin{minipage}{0.4\textwidth}
\includegraphics[keepaspectratio=true,trim = 57mm 140mm 20mm 120mm, clip=true, width=1.0\textwidth]{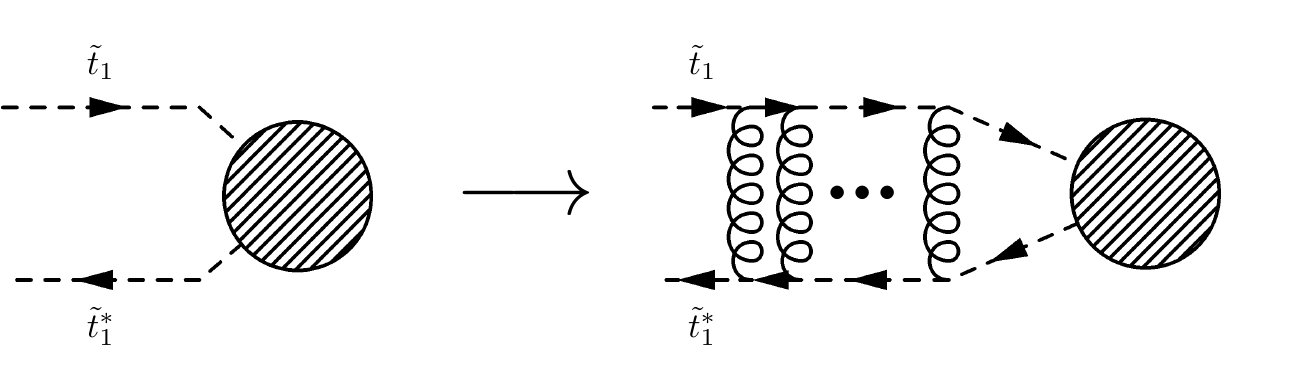}
\end{minipage}
\caption{Left: Tree-level diagrams contributing to stop-antistop annihilation into electroweak SM final states.
	 Hereby, $V=\gamma, Z^0, W^{\pm}$, $H={h^0,H^0,A^0,H^{\pm}}$, and $\ell$ ($\bar{\ell}$) can be any (anti)lepton. Right: Ladder diagram for a leading-order Coulomb potential.}
	\label{Fig:StopStop}
\end{figure*}

For even smaller mass differences between the lightest stop and the lightest neutralino, stop-antistop annhilation becomes the dominant contribution. The relevant tree-level diagrams with leptons, Higgs or electroweak vector bosons in the final state are depicted in Fig.~\ref{Fig:StopStop} (left).
As Coulomb enhancement effects arise in the NLO calculations, these processes have a striking effect on the relic density calculation. The stop-antistop pair is moving slowly during freeze-out, such that  an exchange of $n$ gluons is possible and thus lead to a correction factor proportional to $(\alpha_s / v)^n$, see Fig.~\ref{Fig:StopStop} (right).
With $\alpha_s/v > \mathcal{O}(1)$ under these conditions, the corrections become sizeable, perturbativity breaks down, and for a reliable result, these corrections have to be resummed to all orders within the framework of nonrelativistic QCD. For further details on the calculation, we refer to Ref.~\cite{DMNLO_Stopstop}. We study the impact of the pure NLO calculation and the resummed Coulomb enhancement effects in example scenario C as given in Tab.\ref{Tab:Scenarios}. In this scenario, stop-antistop annihilation contributes with $67.3\%$ to the total cross section. Hereby, the dominant contribution is $\tilde{t}_1 \tilde{t}^*_1 \to h^0 h^0 $ with $46.1\%$, followed by $\tilde{t}_1 \tilde{t}^*_1 \to Z^0 Z^0$ and $\tilde{t}_1 \tilde{t}^*_1 \to W^+ W^-$ with $12.5\%$ and $8.7\%$, respectively.
\begin{figure*}
	\includegraphics[width=0.49\textwidth]{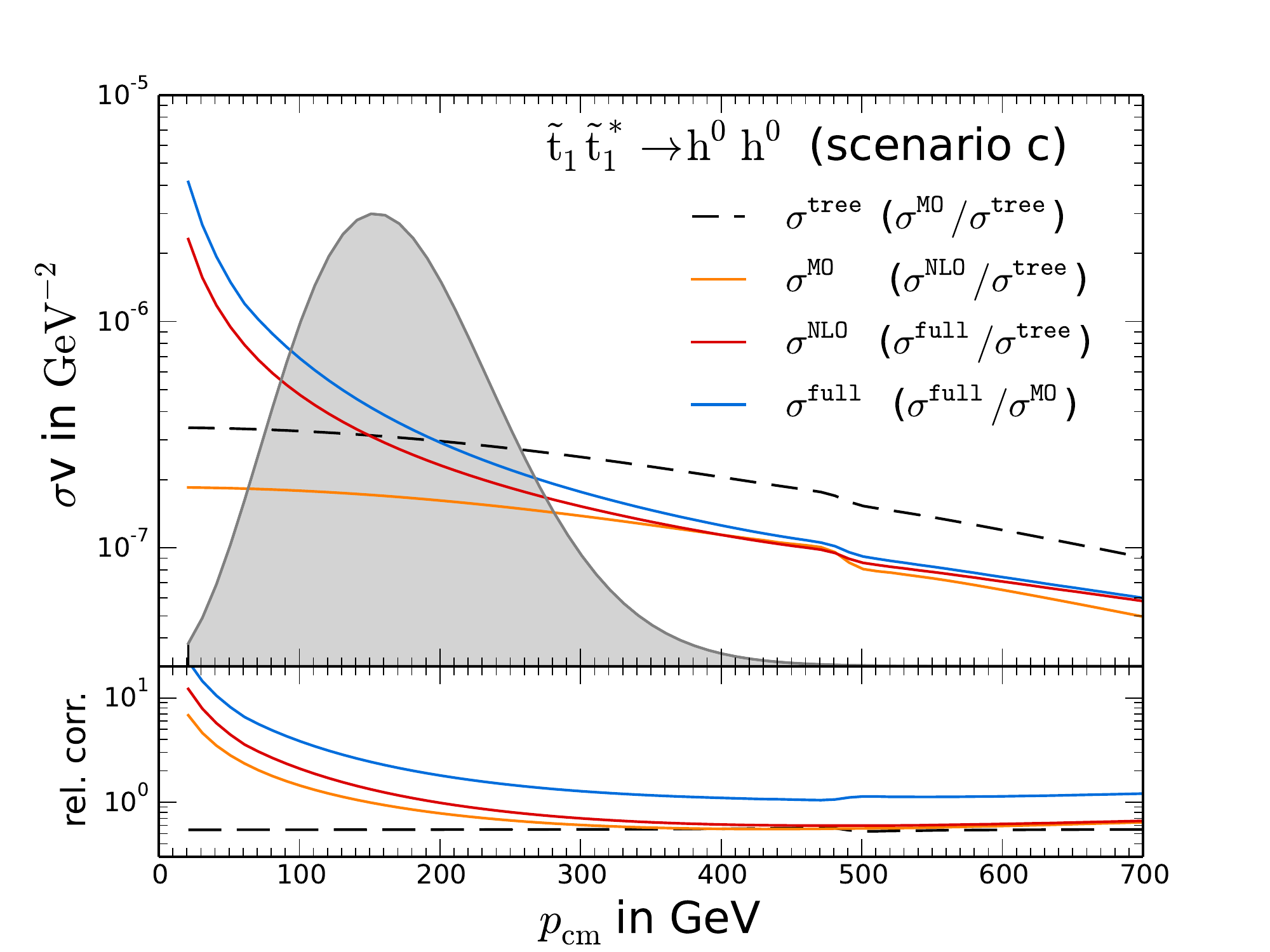}
	\includegraphics[width=0.49\textwidth]{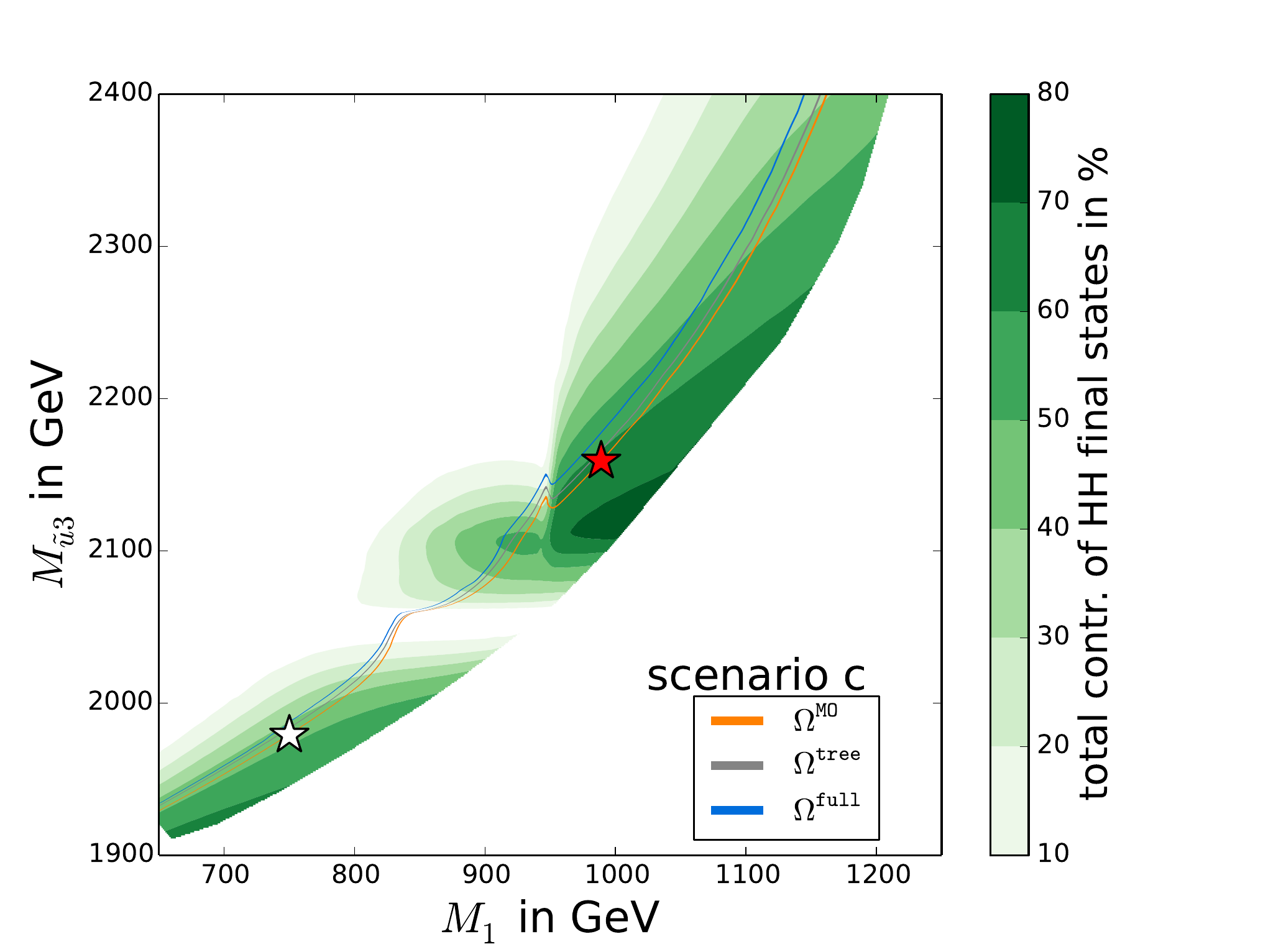}
	\caption{Left: Tree level (black dashed line), {\tt MO} (orange solid line), NLO ($\mathcal{O}(\alpha_s)$) corrections 
	(red solid line) and full corrections (blue solid line) for $\tilde{t}_1 \tilde{t}^*_1 \to h^0 h^0 $ (upper part). The lower part of the plot shows the corresponding 
	ratios of the cross sections. Right: $M_1$--$M_{\tilde{u}_3}$ plane indicating the region of parameter space compatible with Planck, based on {\tt MO} (orange), our tree-level (grey) and 
	our full corrections (blue). The white star marks the example scenario C.}
	\label{Fig:StopStop_results}
\end{figure*}
Fig.~\ref{Fig:StopStop_results} shows the impact of the higher-order corrections, exemplarily for the leading contribution $\tilde{t}_1 \tilde{t}^*_1 \to h^0 h^0 $. Again, there is a deviation between the {\tt MO} and our LO calculation, triggered by the different definition of the top mass and other renormalisation scheme related reasons. In red, the full NLO calculation is depicted. It shows a clear increased cross section for smaller $p_{\mathrm{cm}}$ triggered by the Coulomb enhancement factor. In blue, the result is shown when resumming the Coulomb corrections to all orders. This leads to a relative correction with respect to our tree-level calculation of roughly $300~\%$ in the relevant region. When comparing the full corrected result with the result given by {\tt MO}, we arrive even at a relative correction of $700-800~\%$. For larger values of $p_{\mathrm{cm}}$, however, the Coulomb corrections become less dominant, and the NLO and resummed result converge.

These results have, of course, significant impact on the relic density prediction. This is shown in Fig.~\ref{Fig:StopStop_results} (right plot), where the white star indicates the example scenario C. The 1-$\sigma$ Planck band is depicted in the $M_1 - M_{\tilde{u}_3}$ plane, based on the result of different calculations ({\tt MO} in orange, our tree level in grey, and the full correction in blue). The full correction leads to a shift of roughly $50~\%$ with respect to {\tt MO}. This shows clearly that NLO corrections and the effect of Coulomb corrections have to be taken into account in order to make a reliable theoretical prediction of the relic density.

\begin{figure*}
        \includegraphics[width=0.49\textwidth]{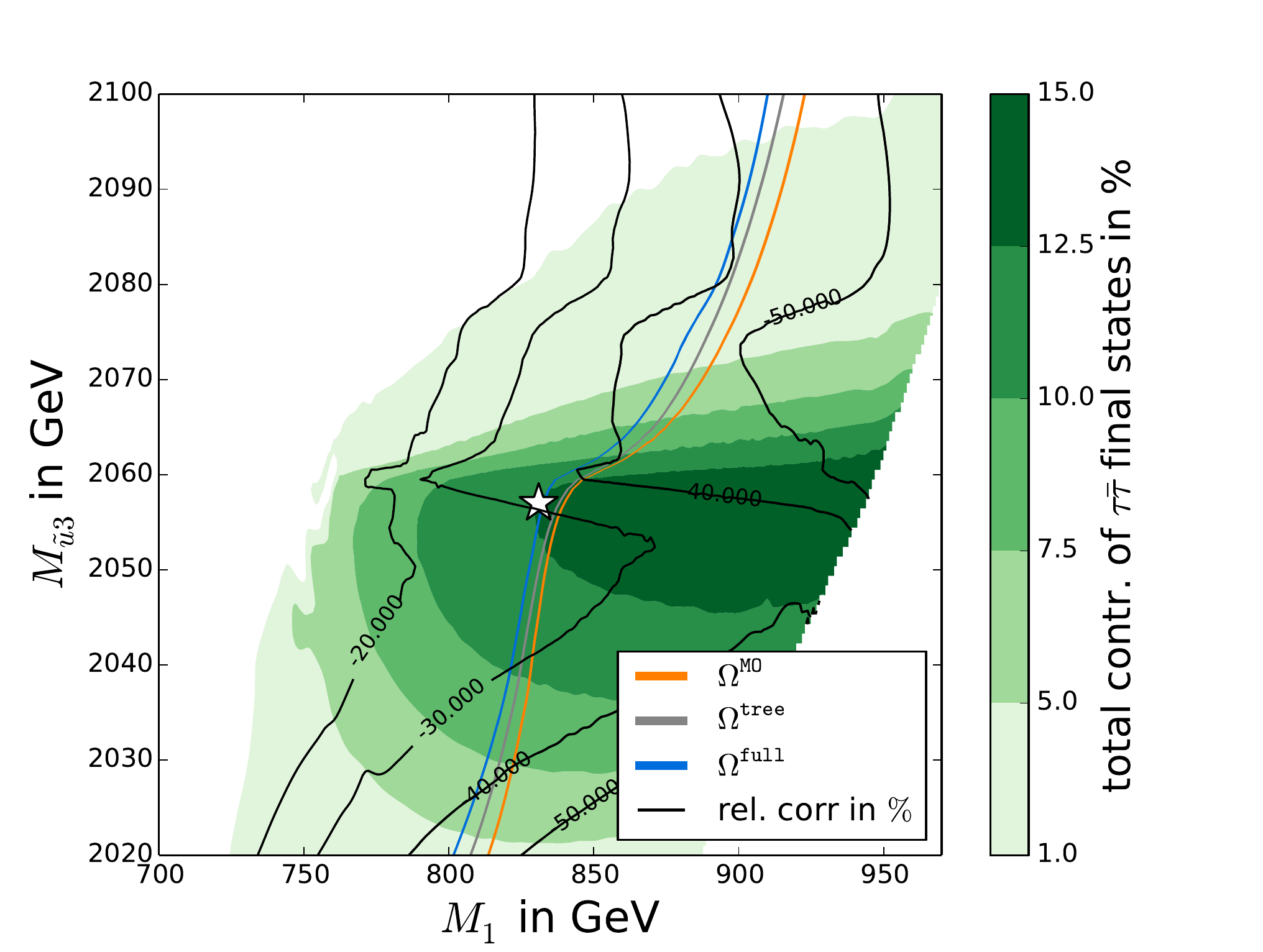}
        \includegraphics[width=0.49\textwidth]{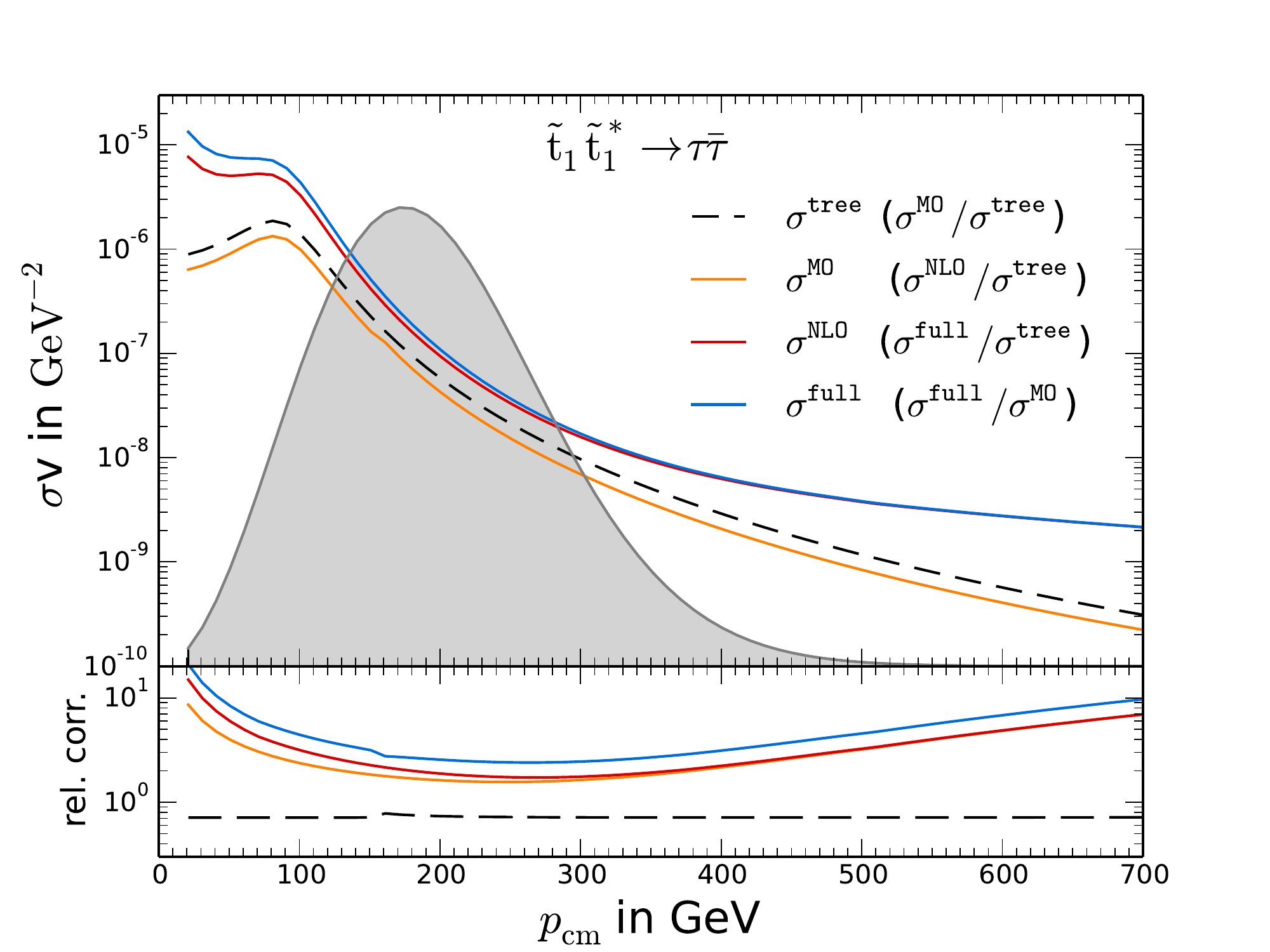}
	\caption{Left: Scan over the $M_1-M_{\tilde{\mathrm{u}}_3}$-plane. The white star marks the position of the new scenario ($M_1=831$ GeV, $M_{\tilde{\mathrm{u}}_3}=2057$ GeV). The black contours depict the deviation between {\tt MO} and our full result. Right: Tree level (black dashed line), {\tt MO} (orange solid line), NLO ($\mathcal{O}(\alpha_s)$) corrections 
	(red solid line) and full corrections (blue solid line) for $\tilde{t}_1 \tilde{t}^*_1 \to \tau \overline{\tau}$.}
	\label{Fig:StopStopAddidum}
\end{figure*}

Another interesting feature concerns the annihilation into a lepton-antilepton pair. Although not dominant in the studied scenario, there is a region in the vicinity of scenario C ($M_1=831$ GeV, $M_{\tilde{\mathrm{u}}_3}=2057$ GeV, marked by a white star in Fig.~\ref{Fig:StopStopAddidum}) where this kind of process contributes to $13\%$. This behaviour is triggered by an s-channel resonance of the heavy Higgs $H^0$, together with a Yukawa coupling which favours for $\tan \beta = 16.3$ the down-type fermions. Although the overall contribution is small, it causes a shift of the relic density of around 20\% with respect to our tree-level calculation and of around 30\% with respect to {\tt micrOMEGAs}. This manifests again that the calculation of SUSY-QCD NLO corrections can be significant and can lead to interesting effects within the MSSM parameter space.

\section{Conclusions}
We have calculated the SUSY-QCD next-to-leading order corrections to gaugino (co)annihilation, neutralino-stop coannihilation, and stop-antistop annihilation into electroweak final states. Depending on the processes, corrections of $10\%$, $18\%$ or $50\%$ including Coulomb resummation, have been observed. As the size of the correction highly depends on the parameter space, it is crucial to take into account those corrections for a precise and reliable theoretical prediction of the neutralino relic density. Thus, it might be interesting to study as well the impact of SUSY-QCD NLO corrections on global fits and other parameter studies within the MSSM.

\section*{Acknowledgements}
The speaker would like to thank the organizers of EPS HEP 2015 for the opportunity to present our work and to contribute to the proceedings. Further, the authors would like to thank Moritz Meinecke for the collaboration on presented projects. The work of JH was supported partly by the London Centre for Terauniverse Studies, using funding from the European Research Council via the Advanced Investigator Grant 267352.

\end{document}